\documentclass[prb,aps]{revtex4}
\usepackage{bm}
\usepackage{graphics}
\usepackage{graphicx}
\usepackage{amscd}
\usepackage{amsmath}
\usepackage{subfigure}
\begin{document}

\title{Voltage Controlled Switching Phenomenon in Disordered
Superconductors}

\author{Sanjib Ghosh,$^{\ast}$ and D. De Munshi$^{\dag}$ }

\affiliation{Center for Quantum Technologies, National University of
Singapore, Singapore 117543}

\begin{abstract}
Superconductor-to-Insulator transition (SIT) is a phenomenon
occurring for highly disordered superconductors and is suitable for
a superconducting switch development. SIT has been demonstrated to
be induced by different external parameters such as temperature,
magnetic field, electric field, etc. However, electric field induced
SIT (ESIT), which has been experimentally demonstrated for some
specific materials, holds the promise of any practical device
development. Here, we demonstrate, from theoretical considerations,
the occurrence of ESIT. We also propose a general switching device
architecture using ESIT and study some of its universal behavior,
such as the effect of sample size, disorder strength and temperature
on the switching action. This work provides a general framework for
development of such a device.
\end{abstract}
\maketitle

\section*{Introduction}

Superconducting switch has been in development for the past 60
years. The first attempt was the development of cryotron, which was
a magnetic field driven switching of a superconductor \cite{Buck}.
There has also been several attempts to generate FET architecture
using superconductors\cite{Nishino},\cite{Schon}.\\
The discovery of superconductor to insulator transition (SIT) for
disordered superconductors \cite{Liu} opened up doors for a new
switching mechanism. SIT was particularly attractive, because unlike
the metal to superconductor transition, SIT provided a much larger
change of the current with one phase being superconductor and hence
zero resistance and the other being insulator and hence infinite
resistance (ideally). However, such transition was driven by either
magnetic field, or disorder modification or temperature change
\cite{Goldman}, making such phenomenon unsuitable for application
in integrated circuits.\\
Quite recently there have been a few demonstrations of electric
field driven SIT. Though these works hold the promise of leading to
a further development of superconducting electronics, all of them
are demonstrated for very specific materials and no microscopic
analysis of such process was given
\cite{Schon},\cite{Gabay},\cite{Parendo} .\\
In this work, we first demonstrate strong fluctuation of
superconducting pair amplitude with electron density (number of
electrons per lattice site), for a strongly disordered
superconductor system. We start with  negative U Hubbard model, Eq.
\ref{hub_mod}, describing a disordered superconductor. From this
model, we then show the strong dependence of the superconducting
pair amplitude, an internal parameter governing the
superconductivity of a sample, on the average density of electron
per lattice site. We then demonstrate that such strong fluctuations
can lead to SIT, through phase correlation calculations. Based on
this phenomenon, we then propose a general architecture of a
superconducting switch, a device, capable of switching from
superconducting state, with effectively zero resistance, to an
insulating state, with resistance of the order of $10 k \Omega$
\cite{Goldman}. Even though there are few realizations of such a
device, most of them requires a large change of electron density to
bring about a change of phase, as is evident from the high values of
voltage needed to switch such a system. The device we are proposing
is driven by a quantum phenomenon
\cite{Ghosh},\cite{Garcia},\cite{Bose}, where small changes of
electron density can lead to a change of phase, hence requiring a
small amount of voltage change, compared to current
experimental devices.\\
Finally, we study some universal properties , namely the effect of
sample size, disorder strength and temperature
on the behaviour of the device.\\

\section*{Model and Methods}
We model the disordered superconductor using a negative-U Hubbard
Hamiltonian on a $L \times L$ square lattice. The Hamiltonian is
given by

\begin{eqnarray}
H=-t\sum_{<ij>,\sigma}(C^{\dag}_{i\sigma}C_{j\sigma}+C^{\dag}_{j\sigma}C_{i\sigma})
+\sum_{i,\sigma}(V_{i}-\mu)C^{\dag}_{i\sigma}C_{i\sigma}\nonumber\\
-U\sum_{i}C^{\dag}_{i\uparrow}C_{i\uparrow}C^{\dag}_{i\downarrow}C_{i\downarrow}.
\label{hub_mod}
\end{eqnarray}
Here $C^{\dag}_{i\sigma}$($C_{i\sigma}$) is the creation
(annihilation) operator for an electron at site $i$ with a spin
$\sigma$, $t$ represents the hopping energy, $V_{i}$ is a site
dependent random potential with uniform distribution from $-V$ to
$+V$, $\mu$ is the chemical potential and $U$ is the strength of the
attractive interaction between two electrons of opposite spins at
the same site.\\
In this model, $t$ represents the kinetic energy of the electrons
and all other parameters are scaled with $t$. $U$ represents the
same site interaction between electrons of the opposite spins and
represent the cooper attraction giving rise to superconductivity.\\

The partition function for this model is given by,
\begin{eqnarray}
 Z &=& \int \mathcal{D}[ C_i,C_i^\dag ] \exp \left(-\int_0^\beta
 \left[ \sum_{i\sigma} C_{i\sigma}^\dag (\tau) (-\partial_\tau+V_i-\mu)C_{i\sigma}(\tau)
 \right.\right. \nonumber \\
 && \left. \left. -t\sum_{<ij>\sigma}(C^{\dag}_{i\sigma}C_{j\sigma} + h.c.)-U\sum_{i}C^{\dag}_{i\uparrow}
 C_{i\uparrow}C^{\dag}_{i\downarrow}C_{i\downarrow} \right] \right)
\end{eqnarray}
where $h.c.$ is the hermitian conjugate and $\beta$ is the inverse
temperature in the unit where Boltzmann constant is unity. We
introduce the Hubbard-Stratonovic transformation with a local
Hubbard-Stratonovic field given by $\Delta_{i}~e^{i\theta_i}$.
$\Delta_{i}e^{i \theta_{i}}$ is the order parameter with
$\Delta_{i}$ being the pair amplitude and $\theta_{i}$ being the
phase. Under this transformation, the partition function becomes,

\begin{eqnarray}
 Z &=& \int \mathcal{D}[ C_i,C_i^\dag ]\mathcal{D}[\Delta_{i}]\mathcal{D}[\theta_{i}] \exp \left(-\int_0^\beta d\tau
 \left[ \sum_{i\sigma} C_{i\sigma}^\dag  (-\partial_\tau+V_i-\tilde{\mu}_i)C_{i\sigma}  \right. \right. \nonumber \\
&& \left. \left.- t\sum_{<ij>\sigma}(C^{\dag}_{i\sigma}C_{j\sigma} + h.c.)-\sum_{i}\Delta_{i}(e^{i\theta_i}
 C^{\dag}_{i\uparrow}C^{\dag}_{i\downarrow}+e^{-i\theta_i}C^{\dag}_{i\downarrow}C_{i\uparrow}) \right] \right)
\end{eqnarray}
where $\tilde{\mu}_i=\mu+\frac{U}{2}n_{i}$ with
$n_{i}=\sum_{\sigma}<C^{\dag}_{i\sigma}C_{i\sigma}>$ which would
be calculated by self-consistency. \\

 {\bf{Bogoliubov-de Genne approximation:}}
The Hubbard-Stratonovic field ($\Delta_{i}~e^{i\theta_i}$) can be
obtained by applying Bogoliubov-de Genne approximation (BdG). In BdG
approximation the partition function is evaluated at the saddle
point. Under this approximation, we obtain an effective Hamiltonian
given by
\begin{eqnarray}
H' &=& -t\sum_{<ij>,\sigma}(C^{\dag}_{i\sigma}C_{j\sigma}+C^{\dag}_{j\sigma}C_{i\sigma})+\sum_{i,\sigma}(V_{i}
-\tilde{\mu}_i)C^{\dag}_{i\sigma}C_{i\sigma} \nonumber \\
&+& \sum_{i}\Delta_{i}(e^{i\theta_i}C^{\dag}_{i\uparrow}C^{\dag}_{i\downarrow}+e^{-i\theta_i}C_{i\downarrow}C_{i\uparrow})
\end{eqnarray}
The Hamiltonian has to follow two self-consistent relations, namely,
$\Delta_{i}e^{i\theta_i}=-U<C_{i\downarrow}C_{i\uparrow}>$
and $n_{i}=\sum_{\sigma}<C_{i\sigma}^\dag C_{i\sigma}>$. $H'$ is
diagonalized by a Bogoliubov transformation
$\gamma_{_{\lambda\sigma}}=\sum_{i}(u_{_\lambda}(i)C^{\dag}_{i\sigma}+\sigma
v_{_\lambda}(i) C_{i\sigma})$. The local Bogoliubov amplitudes are
obtained by the following equation,
\begin{eqnarray}
\begin{pmatrix}
    \hat{{\mathcal{H}}} & \Delta_{i}e^{i\theta_i} \\
    \Delta_{i}e^{-i\theta_i} & -\hat{\mathcal{H}}
\end{pmatrix}
\begin{pmatrix}
u_{_\lambda}\\
v_{_\lambda}
\end{pmatrix}
=E_{_\lambda}
\begin{pmatrix}
u_{_\lambda}\\
v_{_\lambda}
\end{pmatrix}.
\label{bdg.matrix.eq}
\end{eqnarray}
Here $\hat{\mathcal{H}}$ represents the single particle contribution
of $H'$ and $E_{_\lambda}$ are the eigenvalues \cite{Ghosal}. The
self consistent relations in terms of the Bogoliubov amplitudes are
\begin{equation}
\Delta_{i} e^{i\theta_i}=U\sum_{\lambda}u_{_\lambda}(i)v^{*}_{_\lambda}(i)
\end{equation}
\begin{equation}
n_{i}=2\sum_{\lambda}|v_{_\lambda}(i)|^{2}
\end{equation}
Starting from some initial guess values, we self-consistently obtain
the values of $\Delta_{i}e^{i\theta_{i}}$ and $n_{i}$ for each
lattice site $i$. We define the spatial average of $\Delta_{i}$ as
$\Delta_{op}$ given by
\begin{equation}
\Delta_{op}=\frac{1}{N}\sum_{i}\Delta_{i}.
\end{equation}
The average electron density per lattice site is defined as
\begin{equation}
n=\frac{1}{N}\sum_{i}n_{i}.
\end{equation}
We can change the average electron density ($n$) in the sample by
controlling the chemical potential ($\mu$). The BdG approximation
gives the saddle point solution for the $Z$. However, it has
completely missed the fluctuations of phase $\theta_i$, due to
its mean-field nature. \\

{\bf{Fluctuations around the Saddle point:}} To incorporate phase
fluctuations we go beyond BdG and introduce a newly developed method
\cite{Dubi},\cite{Mayr} which allows us to calculate classical phase
fluctuations while ignoring time dependence of the order parameter
(quantum fluctuations). Under this approximation the partition
function becomes,
\begin{equation}
Z=\int\mathcal{D}[\Delta_{i}]\mathcal{D}[\theta_{i}]exp(-\frac{\beta}{U}\sum_{i}|\Delta_{i}|^2)
Tr[ \exp(-\beta H')].
\end{equation}
In terms of eigenvalues of equation(\ref{bdg.matrix.eq}), the
partition function reads,
\begin{equation}
Z=\int\mathcal{D}[\Delta_{i}]\mathcal{D}[\theta_{i}]exp(-\frac{\beta}{U}\sum_{i}|\Delta_{i}|^2)
\prod_{\lambda=1}^{2N}(1+exp(-\beta E_{_\lambda})).
\end{equation}
For obtaining the fluctuations around the saddle point, we relax the
self-consistent constraint on $\{ \Delta_i,\theta_i \}$ and
calculate the value of $E_{\lambda}$ for all possible values of $\{
\Delta_i,\theta_i \}$ through Eq. (\ref{bdg.matrix.eq}) (for a
particular disorder realization).Using the values of $E_{\lambda}$
thus obtained, we can evaluate the values of $Z$ and the expectation
value of any observable $\mathcal{O}$ given by
\begin{eqnarray}
\langle\mathcal{O}\rangle=\frac{1}{Z}\int\mathcal{D}[\Delta_{i}]\mathcal{D}[\theta_{i}]
\mathcal{O}(\theta_{i},\Delta_{i})
exp(-\frac{\beta}{U}\sum_{i}|\Delta_{i}|^2)\nonumber\\
\times\prod_{\lambda=1}^{2N}(1+exp(-\beta E_{_\lambda})).
\label{en_avg}
\end{eqnarray}
We have checked that for the temperature range we are interested in,
the values of $\langle\Delta_i\rangle$ are practically the same as
its value evaluated in BdG approximation. However, $\theta_i$ has a
strong dependence on temperature. Therefore we replace $\Delta_{i}$
in Eq. \ref{en_avg} by its BdG value and hence we only need to
integrate over $\theta_i$. The integration over $\theta_{i}$ is
performed using Monte Carlo method \cite{Dubi},\cite{Mayr}. The
energy eigenvalues
$E_{\lambda} \gg T$ can be ignored while calculating Eq. \ref{en_avg}.\\
Using the partition function $Z$, it has been shown that for weak
disorder, the system shows small fluctuations of pair-amplitudes
while preserving long range phase correlation. Strong disorder on
the other hand, leads to strong fluctuations of the pair-amplitudes
because of the formation of superconducting islands and also
destroys the long range phase correlation. This phenomenon has also
been recently experimentally observed \cite{Pratap}
\cite{Sacepe_prl}.

\section*{Electric field driven SIT (ESIT)}
\begin{figure*}[tbp]
\centering \subfigure[]{\includegraphics[scale=.4]{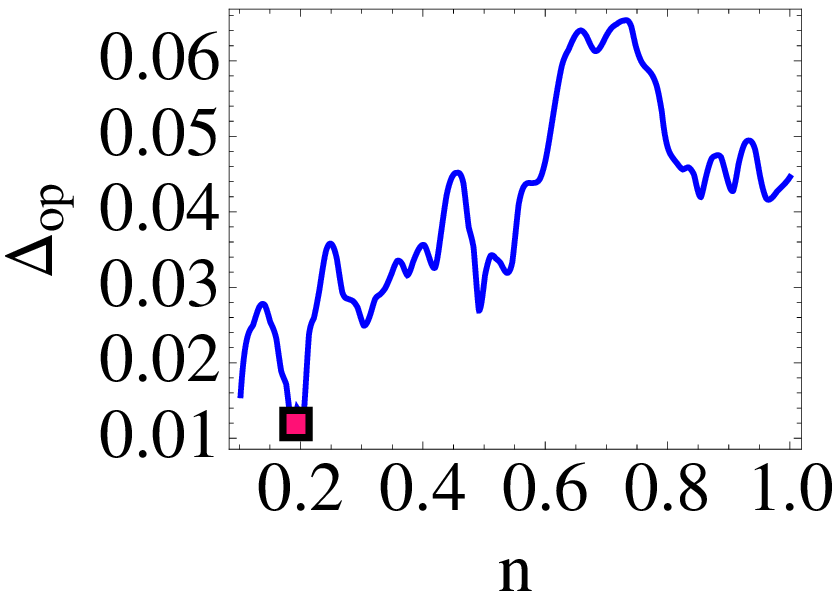}}
\subfigure[]{\includegraphics[scale=.4]{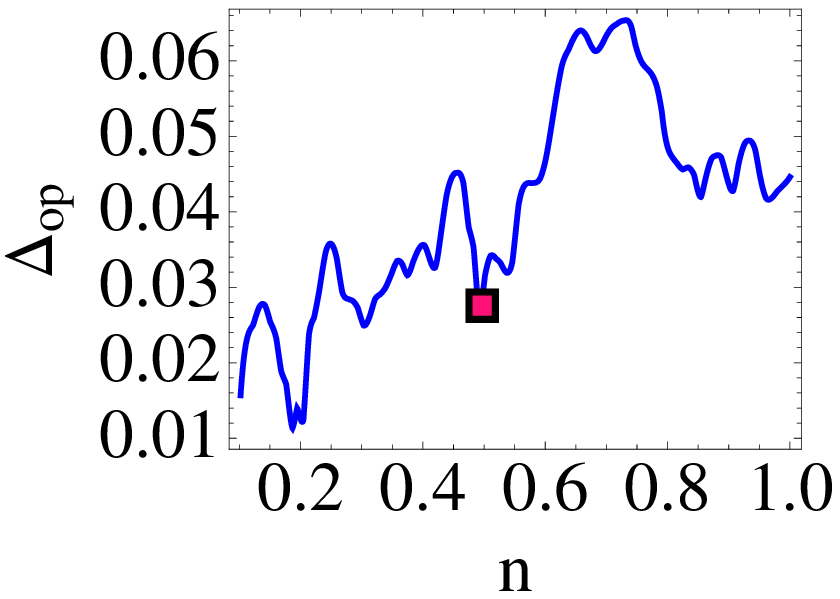}}
\subfigure[]{\includegraphics[scale=.4]{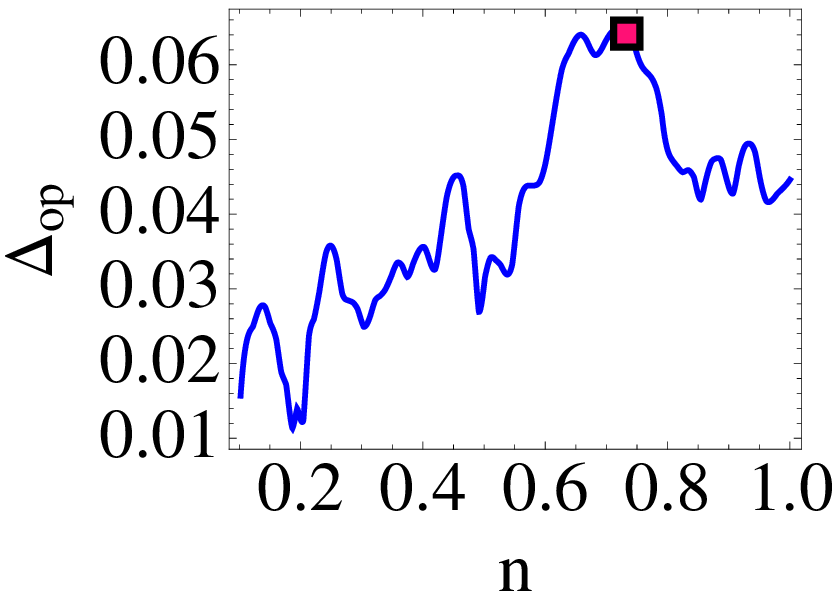}}
\subfigure[]{\includegraphics[scale=.4]{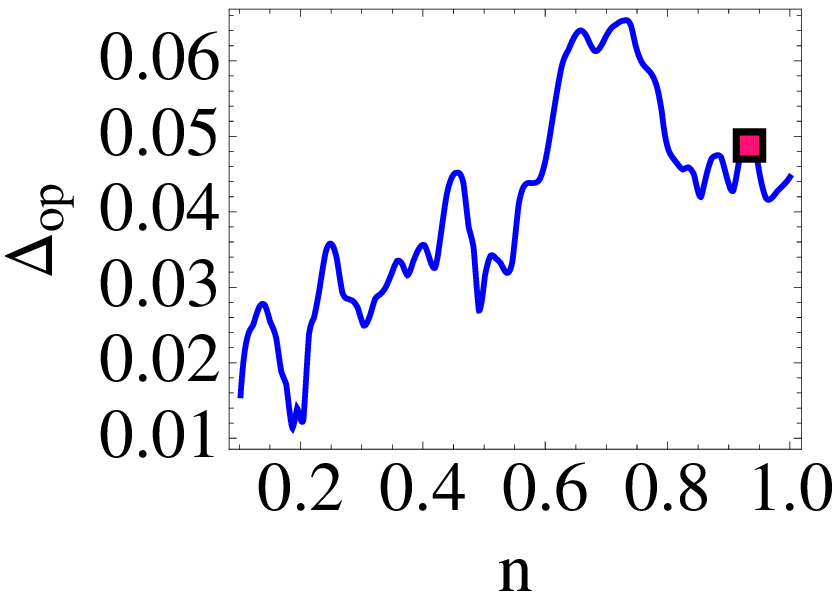}} \\
\subfigure[]{\includegraphics[scale=1.4]{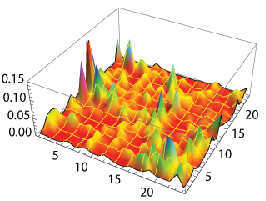}}
\subfigure[]{\includegraphics[scale=1.4]{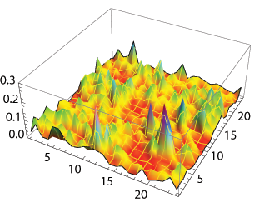}}
\subfigure[]{\includegraphics[scale=1.4]{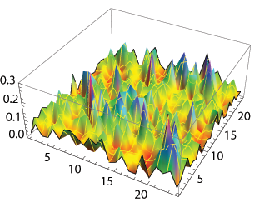}}
\subfigure[]{\includegraphics[scale=1.4]{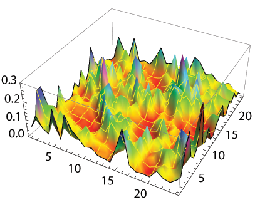}}\\
\caption{Figure demonstrates the dependence of the spatial
distribution of the superconducting order parameter on electron
density for a fixed disorder realization (spatial distribution) and
strength. The plot shows the variation of $\Delta_{op}$ (spatial
average of the order parameter) with electron density per lattice
site ($n$). Figures 1(a),(b),(c) and (d) show the variation of
$\Delta_{op}$ with $n$ for a given disorder realization. The red
boxes denote the value of $\Delta_{op}$, $n$ for which the spatial
distribution of the order parameter is shown through figures
(e),(f),(g) and (h) respectively. The spatial distribution graphs
show that a smaller value of $\Delta_{op}$ corresponds to smaller
and widely spaced superconducting islands ((a),(e) and (b),(f)),
while larger values of $\Delta_{op}$ correspond to larger and
closely spaced superconducting islands ((c),(g) and (d),(h)).
Closely spaced islands allow us to expect that the system for higher
values of $\Delta_{op}$ belong to superconducting phase (g,h) where
as for sparsely spaced islands, corresponding to lower values of
$\Delta_{op}$, we expect a insulating phase(e,f). For switching, we
vary the electron density $n$ such that the system moves from an
insulating state (f, corresponding to n=0.49) to a superconducting
state (g, corresponding to n=.73). The system size used is
$24\times24$ with $V~=~2$ and $U~=~1.5$. } \label{delwn.fig}
\end{figure*}

At strong disorder, $\Delta_{op}$ is strongly dependent on the value
of $n$, as is demonstrated in Figure $\ref{delwn.fig}$. This strong
fluctuation of $\Delta_{op}$ with $n$ is due to rapid change in
local density of states around the small window near the
fermi-surface \cite{Ghosh}. The key feature of this fluctuation is
that, the fluctuation only takes place when the sizes of the
superconducting islands become comparable to the superconducting
coherence length, $\xi_0$. Our main motivation is to induce an SIT
by controlling the size and distribution of the islands, which can
be achieved by tuning $n$.

The tuning of the electron density can be achieved by applying
suitable electric fields. For a first order calculation, we assume a
classical dependence. The electric field inside the two electrodes
separated by a distance $d$ is $V_g/d$, where $V_{g}$ is the applied
voltage. Therefore charge density on the surface is $\epsilon V_g/d$
where $\epsilon$ is the dielectric constant. Because of the
condition of equilibrium in a metal the additional charge density on
the surface of the superconductor is exactly equal to $\epsilon
V_g/d$. To convert this charge density into the electron density per
lattice we divide it by $e/a^2$ where $e$ is electronic charge and
$a$ is the lattice constant on the superconducting plane. Therefore,
we obtain the dependence of electron density on applied electric
field as
\begin{equation}
n(V_{g})=n(0)+\alpha V_{g},
\label{vg.eq}
\end{equation}
where $\alpha=\epsilon a^2 /(e d)$ and $n(0)$ is the electron
density per lattice site for $V_{g}=0$.\\
The electric field is applied perpendicular to the plane of the
superconductor. It is assumed that the disorder superconductor is
only on the $xy$ plane whereas in the $z$ direction it is metal.
This assumption holds for a thin layer of disordered superconductor
such that the thickness in the $z$ axis is much less than the
coherence length and hence it can be treated as a metal along that
axis. For a more realistic situation, the functional dependence of
$n(V_{g})$ will change, but the basic principal of modification of
electronic distribution via electric field remains the same.

Thus we can obtain a SIT driven by applied potential across the
sample. Since this transition is driven by small changes of the
values of $n$, hence small values of applied potential can lead to a
switching of the sample from superconducting state to an insulating
state.


\section*{Device Construction Using ESIT}

The strong dependence of $\Delta_{op}$ on the electron density and
control of the electron density using an electric field opens up the
possibility of developing a voltage control device which switches
between insulator and superconductor states. Because of this strong
dependence, a small change of electron density ($\Delta n \sim 0.1$)
can drive the system from insulator to superconductor and vice
versa. This, in turn,
implies that a small voltage change (from Eq. (\ref{vg.eq})) is required to drive this switching operation.\\
To demonstrate the switching action, we calculate the edge-to-edge
phase correlation for a given sample as a function of electron
density. The edge-to-edge pase correlation is defined as \cite{Erez}

\begin{eqnarray}
D=\frac{1}{Z}\int\mathcal{D}[\Delta_{i}]\mathcal{D}[\theta_{i}](\sum_{m,n}
\cos(\theta_{m}-\theta_{n}))exp(-\frac{\beta}{U}\sum_{i}|\Delta_{i}|^{2})\nonumber\\
\times \prod_{\lambda=1}^{2N}(1+exp(-\beta E_{_\lambda})).
\end{eqnarray}
Here $m$ and $n$ correspond to site indices at the two opposite
edges of the lattice. We have used a $L \times L$ lattice on the xy
plane with the edges from $x=1$ to $x=L$ and $y=1$ to $y=L$. For
calculating the edge-to-edge phase correlation, we sum over all the
lattice sites on the edge. This is because in an actual device,
phase correlation between all the lattice sites on the edge will
contribute. We have assumed a periodic boundary condition along the
x-axis and an open boundary condition along the y-axis. We assume
the current flowing in the y direction and hence we need to measure
the phase correlation along the y-axis.

A non-zero value of edge-to-edge phase correlation implies a
superconducting state and effectively zero resistance current flow
\cite{Erez} . Lack of such correlation even in the presence of
superconducting islands is typical signature of insulating states
associated with SIT
\cite{Sacepe_prl},\cite{Kowal},\cite{Crane},\cite{Sacepe_nat}. Such
states have much higher resistance ($\sim 10k \Omega$) compared to
the superconducting states and a sample in such a state
can effectively work like an open circuit. Figure $2$ demonstrates the switching phenomenon.\\

\begin{figure*}[tbp]
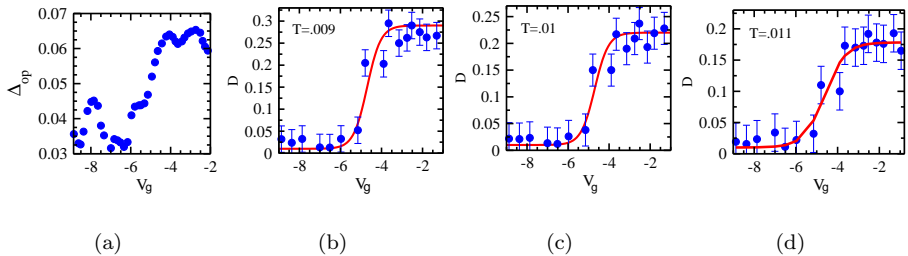

\centering
\subfigure[]{\includegraphics[scale=.43]{switching21.eps}}
\subfigure[]{\includegraphics[scale=.43]{switching22.eps}}
\subfigure[]{\includegraphics[scale=.43]{switching23.eps}}
\subfigure[]{\includegraphics[scale=.43]{switching24.eps}}
\label{switching} \caption{Figure demonstrates the switching
behavior in a region where $\Delta_{op}$ changes significantly. The
edge-to-edge phase correlation ($D$) indicates if the state is
insulating or superconducting. Figure (a) shows the region of
variation of $\Delta_{op}$ with $V_{g}$ (unit=Volts) (through Eq.
\ref{vg.eq}) where we demonstrate the switching phenomenon. This
region is chosen because of large change of $\Delta_{op}$ for a
small change of $n$. The temperature dependence of the switching
process is demonstrated across figures (b),(c) and (d). Increase of
temperature $T$ (scaled with the critical temperature) leads to a
decrease of the gap between the switching states. It should be noted
that for the temperature regime we are working in, the temperature
dependence of $\Delta_{op}$ is insignificant. This demonstration of
switching is for the same sample (same disorder realization and
parameter values) as that for Figure $1$. The change in density due
to the applied voltage is from $0.42 (V_g=-8.5)$ to $0.78
(V_g=-2.0)$, the parameters used in Equation(\ref{vg.eq}) are,
$n(0)=0.87$ and $\alpha=0.053$.}
\end{figure*}

We can now use this switching phenomenon to construct a device which
can act as a voltage controlled electronic switch. The basic
architecture is shown in the Figure $3$. The switching takes place
on the $z=0$
plane whereas the control field acts in the $z$ direction.\\

\begin{figure}[t]
\centering
\includegraphics[scale=.4]{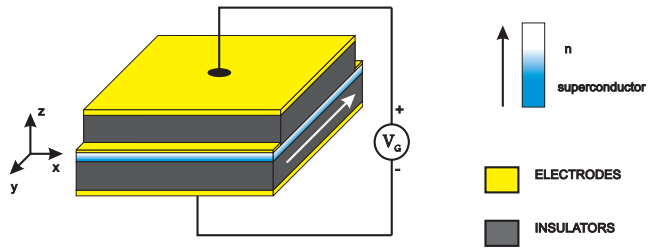}
\label{device} \caption{A switching device using ESIT. The electric
potential applied across the insulators in the z direction ($V_{g}$)
controls the electron density. The origin is on the upper surface of
the superconductor. The current flows on the $z=0$ plane and along
the $y$ axis. This is an idealistic architecture. Architecture for
real implementation might be very different and can depend strongly
on the chosen materials.}
\end{figure}

\section*{Effect of Sample Size, Disorder Strength and Temperature on
Switching}

The operational efficiency of the device depends  on the strength of
the disorder ($V$), the sample size of the device, compared to the
coherence length ($L/\xi_{_{0}}$), and the operating temperature ($T$).\\

{\bf{Effect of Sample Size:}} Figure $4$ demonstrates the effect of
sample size on the fluctuation of the superconducting pair
amplitude. If $L/\xi_{_{0}}\sim 1$, then the fluctuations of
$\Delta_{op}$ with $n$ would increase and the stability of the
switch would be affected. On the other hand, for $L/\xi_{_{0}}
\rightarrow \infty$, the fluctuation reduces and the switching
property can be suppressed. The switching property arises because of
the strong fluctuation of $\Delta_{op}$ with $n$. As shown in Fig.
1, for a particular value of $n$, we have large and closely spaced
superconducting islands, corresponding to large value of
$\Delta_{op}$, whereas, for another value of $n$ we have large
non-superconducting regions, corresponding to smaller values of
$\Delta_{op}$. This is true for a system size comparable to the
coherence length. However, for $L/\xi_{0}\gg1$, even though in
regions of size comparable to $\xi_{0}$, we have strong pair
amplitude fluctuation with $n$, on the scale of the system size,
change of electron density merely rearranges the position of the
superconducting islands and hence the global properties like
$\Delta_{op}$ and edge-to-edge phase correlation doesn't show a
significant dependence on $n$, leading to the suppression of  the
switching property. Thus for efficient operation, suitable sample
size must be selected, depending on the coherence length of
the material used.\\

\begin{figure}[t]
\begin{centering}
\subfigure[]{\includegraphics[scale=.35]{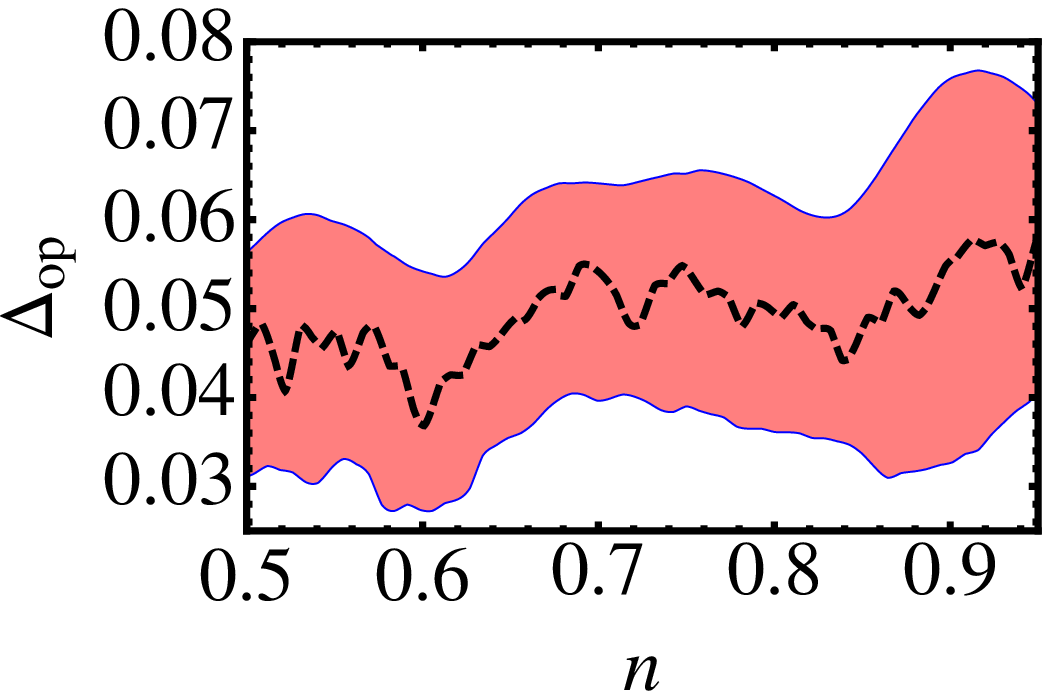}}
\subfigure[]{\includegraphics[scale=.456]{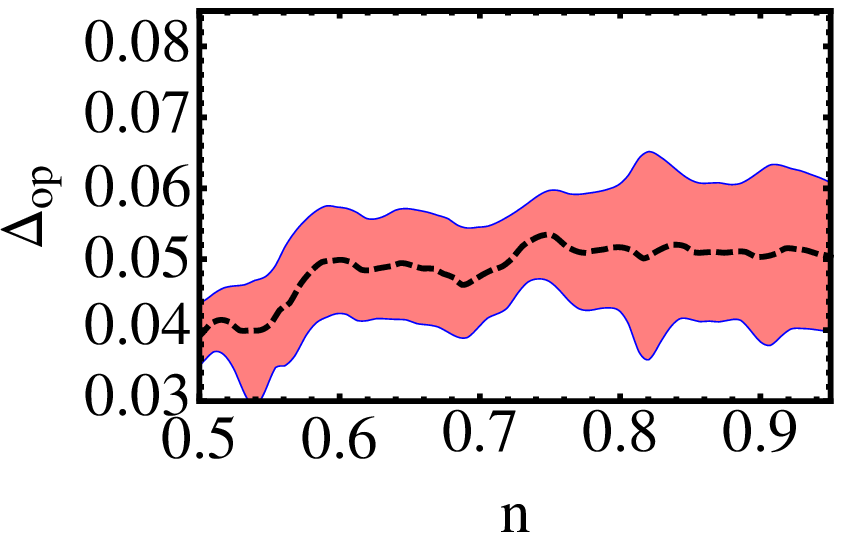}}
\subfigure[]{\includegraphics[scale=.35]{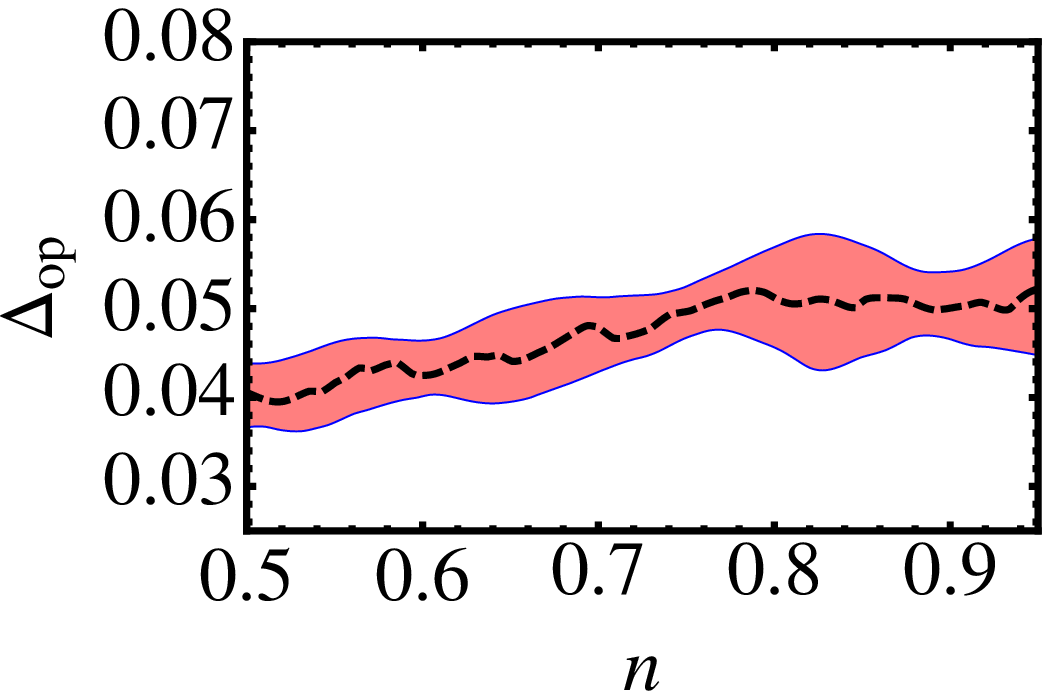}}
\label{sizevswtich} \caption{The figure investigates the dependence
of the behavior of $\Delta_{op}$ with $n$ on the system size. The
red region is the total region of variation of $\Delta_{op}$ with
$n$ for different disorder realizations. The black curve is the
average over different disorder realizations ($5$ to $10$). However,
the disorder averaging is less informative in this context because
we are using small system size (few $\xi_{0}$) with a fixed disorder
landscape. (a),(b) and (c) correspond to $10\times 10$,~$20\times
20$ and $30\times 30$ system sizes. The coherence length
$\xi_{_{0}}$ is of the order $10$. The plots demonstrate that as
$L/\xi_{_{0}}$ increases, the fluctuations decrease. For
$L/\xi_{_{0}}\sim 1$, high fluctuations can make the switching
phenomenon unstable. $L/\xi_{_{0}} \gg 1$ leads to suppression of
the fluctuations and can increase the driving range to obtain the
switching process or suppress it completely. Hence, given a
$\xi_{_{0}}$, an optimum sample size has to be selected for
efficient performance of the device. $V=2$ and $U=1.5$.}
\end{centering}
\end{figure}

{\bf{Effect of Disorder Strength:}} In the weak disorder regime, the
effect of electron density $n$ on the value of $\Delta_{op}$ is
insignificant. However, stronger the disorder, greater is this
effect, as is demonstrated in Fig. 5. This happens because of rapid
change of superconducting landscape of a sample with a small change
of electron density in strong disorder regime. A strong change of
local density of states depending on the value of $n$ leads to this
changing landscape as is described in Ref. \cite{Ghosh}.\\
Sample size and disorder strength give us two handles to control the
change of electron density required for performing the switching
action. By changing the sample size and the disorder strength, we
can change the fluctuation of $\Delta_{op}$ with $n$ and hence we
can control the change of density required for the system to switch
from insulating state to superconducting state and vice versa. For
example, if it is needed to decrease the change of $n$ required for
the switching, one can increase the disorder strength or reduce the
sample size or both.\\


\begin{figure}[t]
\begin{centering}
\subfigure[]{\includegraphics[scale=.45]{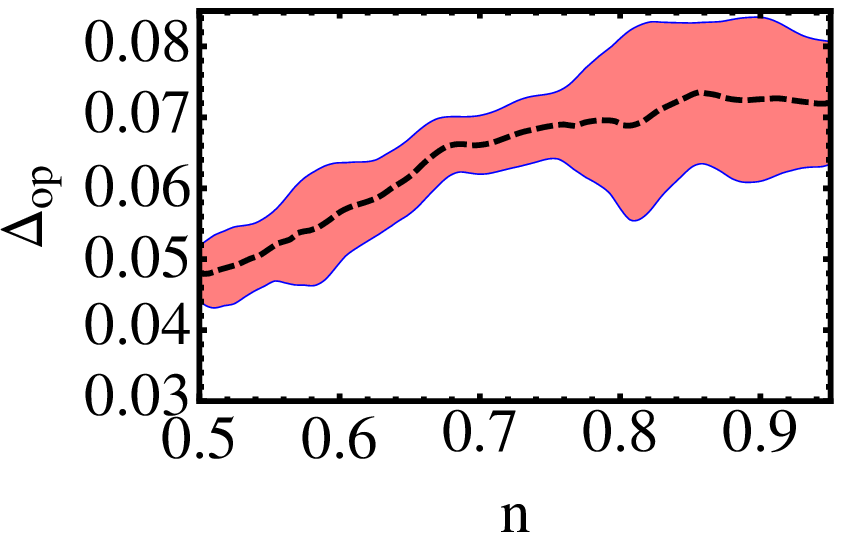}}
\subfigure[]{\includegraphics[scale=.45]{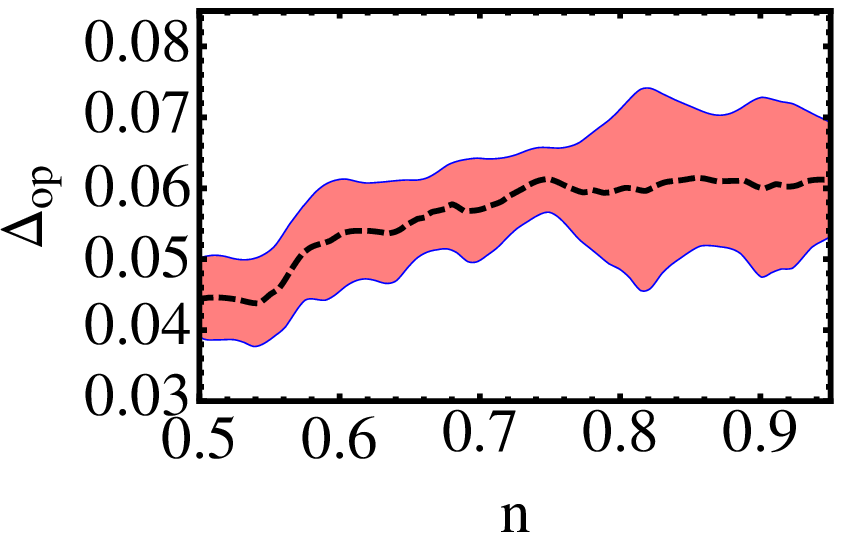}}
\subfigure[]{\includegraphics[scale=.45]{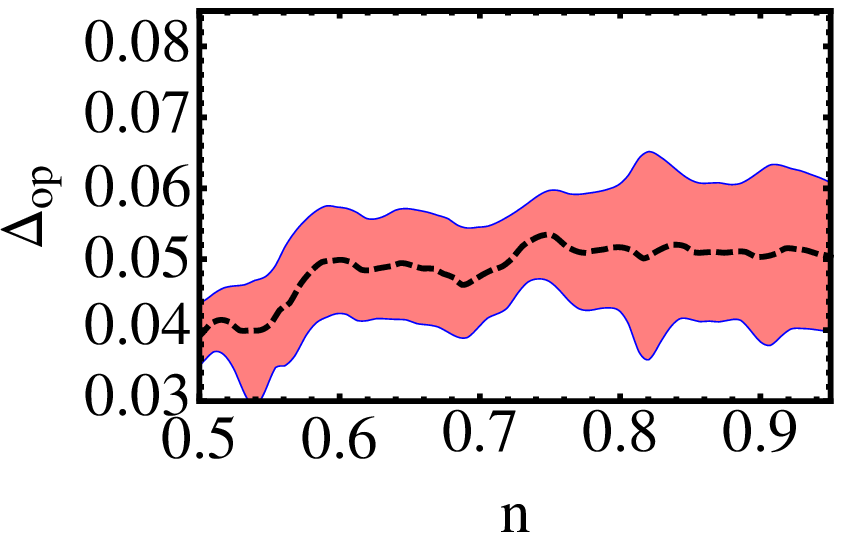}} \caption{Figure
demonstrating the effect of disorder strength on the behavior of
$\Delta_{op}$ with $n$. The system size is kept at $20 \times 20$
for each plot. The red area shows the total region of variation of
$\Delta_{op}$ with $n$ for different disorder realizations. The
black curve is the average over different disorder
realizations($10$). (a),(b) and (c) corresponds to disorder
strengths $V=1.5$, $V=1.75$ and $V=2$. As disorder strength
increases, the fluctuations also increase. The phenomenon behind
this is similar to the dependence on size. Higher disorder strengths
lead to decrease of average size of superconducting islands and
hence increasing the fluctuations as suppression of the switching
behavior. $U=1.5$.}
\end{centering}
\label{sizevswtich}
\end{figure}

{\bf{Effect of Temperature:}} Within the range we are working,
finite temperature will have insignificant effect on the
superconducting landscape and hence $\Delta_{op}$. However, the
edge-to-edge phase correlation will have a strong dependence on the
temperature. This in turn, will effect certain properties of the
switching action, such as the switching point (the voltage or the
electron density about which the system undergoes a transition from
insulator to superconductor), switching gap (the difference of the
value of $D$ across the switching point),etc. As can be seen in
Figure $2$, with change of temperature, the difference between the
switching states changes. This, in turn, can effect the stability of
the switch. Figure $6$ demonstrates the effect of temperature on the
switching gap. However, the switching point ($V_{c}$), i.e.
$D(V_{c})=(D_{UP}-D_{DOWN})/2$, remains independent of temperature,
where $D_{UP}$ and $D_{DOWN}$ are the values of $D$ at
superconducting and insulating states respectively. The invariance
of the switching point with temperature can be attributed to the
fact that for low temperatures, the landscape of pair amplitude
across the sample, remains almost invariant. On the other hand, the
change of the switching gap arises because of the decrease of the
phase correlation in the superconducting state with
an increase of temperature.\\

\begin{figure}[h]
\centering
\includegraphics[scale=.3]{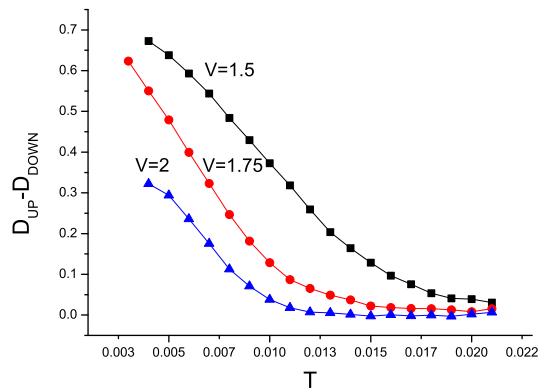}
\label{switchvstemp} \caption{Figure explores the role of
temperature (T) on the behavior of the device. Increase of
temperature reduces the gap in the value of $D$ (hence conductivity)
between the two states. The variation with $T$ also depends on the
strength of the disorder.}
\end{figure}

\section*{Discussions and Conclusions}
The strong dependence of pair amplitude on the average electron
density have thus enabled us to postulate a device, capable of
switching between insulator and superconductor states, driven by
very small change of electron density. The requirement of small
change of electron density in turn implies that a small voltage
change is required to drive the switching mechanism. In the example
we have shown (Fig. $2$), the total change of the electron density
is $0.34$ along the x-axis. However, the change of electron density
across the transition point is as low as $0.1$.\\
We have used the following values of material properties for Eq.
\ref{vg.eq}- $a=3~\text{\AA}$, $d=1\mu m$ and
$\epsilon=3\times10^{4}~\epsilon_{0}$, where $\epsilon_{0}$ is the
free space permittivity. For these parameters, we need a voltage
change of $2~V$ (density change $=~0.1$) for switching. However, by
controlling the disorder and sample size, we can modify the
switching voltage
appropriately.\\
A typical coherence length is of the order of $50~nm$. Therefore a
typical system size can be be of the order of $0.1 \mu m$. However,
one can change the size of a typical device by using materials
having different coherence lengths. For example, by using aluminium,
one can construct devices with size of the order of tens of microns,
where as by using materials such as alloys of $Nb$ and $Sn$, one can
construct devices of the order of tens of nanometers. Also the
architecture provided is a very basic FET structure. But in the a
realistic situation, the architecture might be completely different
and material dependent.\\
Individual characteristics of a device, such as the change in
electron density for the transition, the electron density about
which the transition occurs, the switching gap, etc, are strongly
sample dependent. Different samples with different disorder
realization will have different switching points, switching gaps,
etc. However, the phenomenon of the strong dependence of
$\Delta_{op}$ on $n$, and ESIT is independent of the disorder
landscape. Averaging over different disorder realization will erase
the fluctuations but since a single sample will contain a particular
disorder realization, disorder averaging is less informative in this
context.\\
The architecture of this device gives us an distinct advantage over
the previous attempts, since it can provide a possible adaptation of
superconducting switches in integrated circuits. Also, the
theoretical treatment allows us to claim in a generalized manner
that such a switch can be developed, though the exact material,
optimal for application, can only be determined experimentally. With
proper system, such a design can potentially usher in
superconducting electronics, which can improve
the efficiency and capability of large computational systems.\\

\section*{Acknowledgements}
We would like to thank Prof. Shudhansu Shekhar Mondal for valuable
discussions.


\end{document}